\documentclass[a4paper]{jpconf}
\usepackage[symbol]{footmisc}
\usepackage{graphicx,amsmath,amssymb,floatrow}

\newcommand{\figref}[1]{Fig.~\ref{#1}}
\newcommand{\tabref}[1]{Table~\ref{#1}}
\newcommand{\refref}[1]{Ref.~\cite{#1}}

\makeatletter
\def\blfootnote{\gdef\@thefnmark{}\@footnotetext}
\makeatother

\begin{document}

\title{Precision measurements of $\theta_{12}$ for testing models of discrete leptonic flavour symmetries}

\author{P.~Ballett$^1$$^\dagger$, S.~F.~King$^2$, C.~Luhn$^3$, S.~Pascoli$^1$ and M.~A.~Schmidt$^4$}
\address{$^1$IPPP, Department of Physics, Durham University, South Road, Durham DH1 3LE, United Kingdom}
\address{$^2$School of Physics and Astronomy, University of Southampton, Highfield, Southampton SO17 1BJ, United Kingdom}
\address{$^3$Theoretische Physik 1, Naturwissenschaftlich-Technische Fakult\"{a}t, Universit\"{a}t Siegen, Walter-Flex-Stra\ss{}e 3, D-57068 Siegen, Germany}
\address{$^4$ARC Centre of Excellence for Particle Physics at the Terascale, School of Physics, The University of Melbourne, Victoria 3010, Australia}

\begin{abstract} 
Models of leptonic flavour with discrete symmetries can provide an attractive
explanation of the pattern of elements found in the leptonic mixing matrix. The
next generation of neutrino oscillation experiments will allow the mixing
parameters to be tested to a new level of precision, crucially measuring the CP
violating phase $\delta$ for the first time. In this contribution, we present
results of a systematic survey of the predictions of a class of models based on
residual discrete symmetries and the prospects for excluding such models at
medium- and long-term oscillation experiments. We place particular emphasis on
the complementary role that a future circa~$50$~km reactor experiment,
\emph{e.g.} JUNO, can play in constraining these models.
\blfootnote{\hspace{-0.15cm}$^\dagger$ Poster presenter at NuPhys 2013 held at
Institute of Physics, London, UK.}

\end{abstract}

\section{\label{sec:intro}Introduction}

The next generation of neutrino oscillation experiments has the potential to
not only determine the remaining unknowns in the PMNS matrix but also to
measure its parameters with unprecedented precision. This will mark the beginning
of a period of high-precision neutrino physics, where the standard paradigms
describing the neutrino sector will be put to proof and theoretical ideas about
the origins of neutrino mass and leptonic flavour can be confronted with data.

One of the more popular beyond the standard model ideas applied to the neutrino
sector is the introduction of a discrete flavour symmetry. Models based on this
principle have been shown to be able to derive the observed structure of the
PMNS matrix from a small set of assumptions.  These models generally propose a
discrete symmetry (\emph{e.g.}  $\text{A}_4$ or $\text{S}_4$) which is broken
spontaneously, leaving residual symmetries amongst the leptonic mass terms.
These symmetries reduce the degrees of freedom amongst the mixing parameters,
generating a pattern of falsifiable predictions. 
By hypothesizing which symmetries of the leptonic mass terms are residual, this
idea can be used to reconstruct the flavour group independently of many model
specific assumptions.  In \refref{Ballett:2013wya}, we have shown that a quite
general construction of this type (first presented in
\refref{Hernandez:2012ra}) leads to only $8$ viable models
in light of the current global oscillation data. These models fix a column of
the PMNS matrix which, under the assumption of unitarity, can be expressed in
terms of two constraints on the PMNS parameters: the \emph{atmospheric sum
rule}, relating $\theta_{23}$ to $\theta_{13}$ and $\delta$, and the
\emph{solar prediction}, an expression for $\theta_{12}$ in terms of
$\theta_{13}$ alone. 
In this contribution, we shall discuss the parameter correlations of these
models, and how they can be constrained by the next generation of
high-precision oscillation experiments.
We shall employ the notation $s=\sqrt{3}\sin\theta_{12}-1$,
$r=\sqrt{2}\sin\theta_{13}$ and $a=\sqrt{2}\sin\theta_{23}-1$
\cite{King:2007pr} throughout.

\begin{table}[t]
\begin{tabular}{c c c} Model label & solar prediction & predicted $\theta_{12}$
($r$ in $3\sigma$) \\ \hline\hline A$_4$ T$_\alpha$--S$_2$ & $s
=\sqrt{\frac{2}{2-r^2}} - 1$ & [35.62, 35.86] \\ \hline S$_4$ T$_e$--S$_1$ & $s
= \sqrt{1-\frac{2r^2}{2-r^2}} - 1$ & [34.05, 34.55]\\ S$_4$ T$_\alpha$--S$_2$ &
$s = \sqrt{\frac{3}{2(1-r^2)}} - 1$ & [30.29, 30.49]\\ \hline A$_5$
T$_e$--S$_1$ & $s = \sqrt{ 3 + \frac{6}{(3-\varphi)(r^2-2)}} -1 $ & [30.33,
30.90] \\ A$_5$ T$_e$--S$_2$ & $s = \sqrt{\frac{6}{(2+\varphi)(2-r^2)}} -1$ &
[32.03, 32.24]\\ A$_5$ T$_\alpha$--S$_2$ & $s =
\sqrt{\frac{3\varphi}{(2\varphi-1)(2-r^2)}} -1$ & [37.56, 37.62] \\ \hline
\end{tabular}
\caption{\label{tab:solar_pred}The solar predictions for the $8$ viable models
identified in \refref{Ballett:2013wya}. The model label denotes the flavour
group and the pattern of breaking; for details, see \refref{Ballett:2013wya}.}
\end{table}

\section{\label{sec:atm}Atmospheric sum rules}

The atmospheric sum rule can be written in a linearized form by
\[ a = a_0 + \lambda r \cos\delta  + \mathcal{O}\!\left(r^2\right), \]
where $a_0$ and $\lambda$ are constants expressible in terms of the group
theoretic parameters of each model. To test these relations, we require a
strong precision on the parameter $\delta$, which necessitates the
consideration of the next generation of long-baseline experiments. These
proposals seek to make accurate measurements of the appearance channels
$\nu_\mu \to \nu_e$ and $\overline{\nu}_\mu \to \overline{\nu}_e$, which are
sensitive to the value of $\delta$ at a subdominant level. Although it remains
a challenging measurement, two leading designs have been shown to offer
significant sensitivity to $\delta$: superbeams and neutrino factories. Such
facilities would be able to constrain the atmospheric sum rules over a
significant fraction of the available parameter space. For example, an on-axis
superbeam with a detector mass of $70$~kton ($35$~kton) and a baseline of
$2000$~km would be capable of excluding models with $a_0=0$ and $\lambda=1$ for
over $44$--$88\%$ ($19$--$75\%$) of the parameter space, depending on the true
value of $\theta_{23}$ \cite{Ballett:2013wya}. 

\section{\label{sec:sol}Solar predictions and reactor experiments}

\begin{figure}[t]
\includegraphics[width=9.5cm,clip,trim=3 5 12 5]{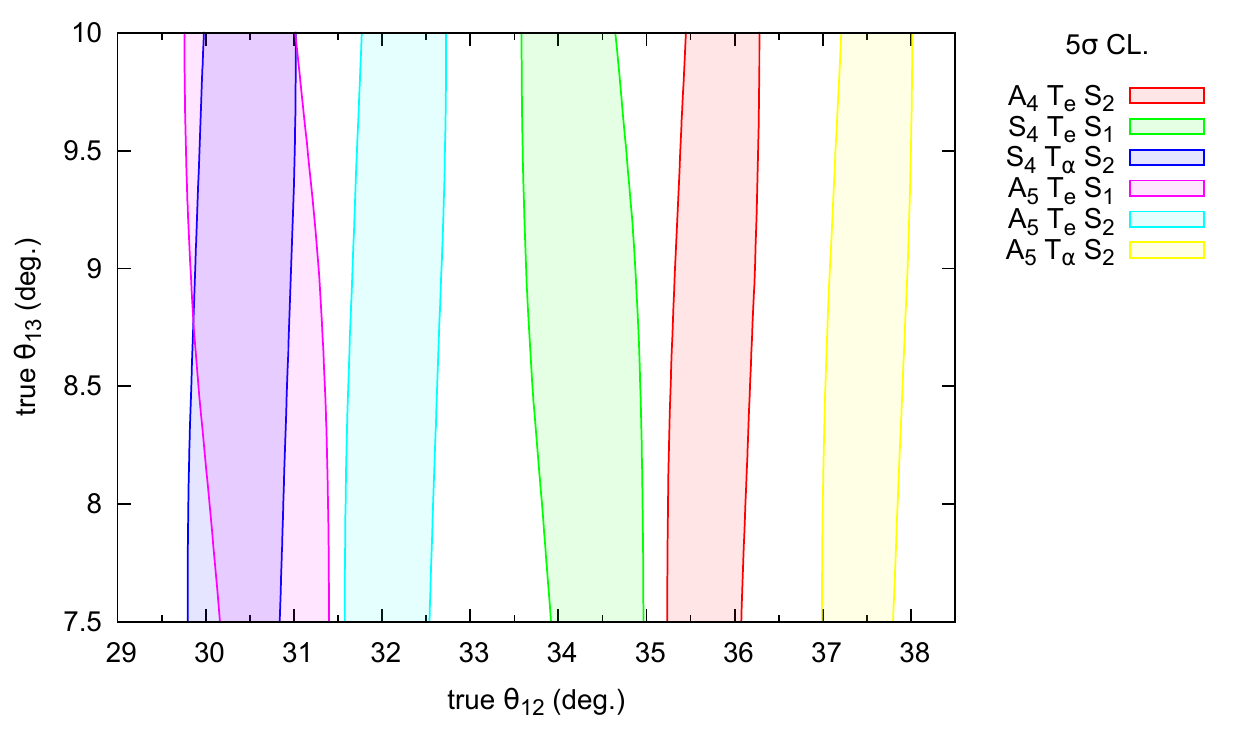}
\caption{\label{fig:JUNO_SR}The $5\sigma$ allowed regions for the solar
predictions shown in \tabref{tab:solar_pred} after $6$ years of data taking by
JUNO.}
\end{figure}

In this section, we shall consider a circa $50$~km reactor experiment based on
the Jiangmen Underground Neutrino Observatory (JUNO) \cite{Li:2013zyd} and
Reactor Experiment for Neutrino Oscillations (RENO-50) designs
\cite{Kim:2013aa}. These facilities will be capable of high precision
measurements of the $\overline{\nu}_e$ disappearance probability. The main goal
of such experiments is to observe the subdominant oscillations whose phase
depends upon the mass hierarchy. However, they will also significantly increase
the precision on the oscillation parameters $\theta_{12}$, $\Delta m^2_{21}$
and $\Delta m^2_{31}$, reducing their uncertainty to the sub-percent level. The
$8$ viable models identified in \refref{Ballett:2013wya} make $6$ distinct
solar predictions, which are shown in \tabref{tab:solar_pred}. Such precision
will have a significant impact on the viability of the correlations predicted
by flavour symmetric models.
To understand the impact of these high precision measurements, we have
performed a simulation based on the JUNO design to determine its ability to
test the correlations shown in \tabref{tab:solar_pred}. In our simulation, we
assume a $20$~kton liquid scintillator detector with a linear energy
uncertainty of $0.03/\sqrt{E}$. The JUNO facility will detect neutrinos from
$10$ nearby reactors; however, we model this by a single source at a baseline
distance given by the power weighted average of $52.5$~km and a reactor power
of $36$~GW \cite{Li:2013zyd}.  We have normalised our spectrum to produce
$10^5$ events, including a $5\%$ normalisation uncertainty. In
\figref{fig:JUNO_SR}, we show the allowed regions at $5\sigma$ significance for
the models shown in \tabref{tab:solar_pred}. We see that only two of the
$5\sigma$ intervals overlap, which allows for a strong model discrimination.
The ability for JUNO to exclude these models independently of their atmospheric
sum rules provides a great complementarity between the reactor and
long-baseline programmes.  Furthermore, the two indistinguishable models for
JUNO predict very different atmospheric sum rules,
\begin{align*} a = \pm\frac{1}{6} - \frac{1}{\sqrt{6}}r\cos\delta~~\text{(S$_4$
T$_\alpha$--S$_2$)}\qquad\text{and}\qquad a =
\frac{\varphi}{\sqrt{2}}r\cos\delta~~\text{(A$_5$ T$_e$--S$_1$)}, \end{align*}
where $\varphi = \frac{1+\sqrt{5}}{2}$ is the golden ratio, and we expect these
to be distinguishable with a superbeam for most of the parameter space
\cite{Ballett:2013wya}. 

\section{Summary}

The next generation of neutrino oscillation experiments, with their focus on
precision measurements of the underlying parameters, will allow certain classes
of models with discrete flavour symmetries to be thoroughly tested. In
\refref{Ballett:2013wya}, the role of a long-baseline superbeam experiment
(modelled after LBNO or LBNE) has been shown to be able to exclude these
correlations for a large fraction of parameter space. In this contribution, we
have highlighted the potential for experimental exclusion of these models at a
circa $50$~km reactor experiment based on the JUNO facility. By testing the solar
predictions to high accuracy, such a facility will be able to independently
distinguish between almost all models under consideration. The complementarity
between reactor and long-baseline experiments will provide a stringent test of
the idea that residual symmetries are responsible for the structure of the PMNS
matrix.


\section*{References}
\begin{thebibliography}{9}

\bibitem{Ballett:2013wya}
  P.~Ballett, S.~F.~King, C.~Luhn, S.~Pascoli and M.~A.~Schmidt,
  Phys.\ Rev.\ D {\bf 89} (2014) 016016.

\bibitem{King:2007pr}
  S.~F.~King,
  Phys.\ Lett.\ B {\bf 659} (2008) 244.

\bibitem{Hernandez:2012ra}
  D.~Hernandez and A.~Y.~Smirnov,
  Phys.\ Rev.\ D {\bf 86} (2012) 053014.

\bibitem{Li:2013zyd}
  Y.-F.~Li, J.~Cao, Y.~Wang and L.~Zhan,
  Phys.\ Rev.\ D {\bf 88} (2013) 1,  013008.

\bibitem{Kim:2013aa}
  S.-B.~Kim,
  talk given at SISAC: from Particle to Universe, SKKU. December 2-4, 2013.

\end{thebibliography}
\end{document}